\pdfoutput=1
\documentclass[12pt,a4paper]{article}
\usepackage{lineno}  

\usepackage{graphicx}  

\usepackage{xspace}
\usepackage{color}
\usepackage{colortbl}
\usepackage{subfigure}
\usepackage{amsmath}

\usepackage{ifthen} 
\newboolean{pdflatex}
\setboolean{pdflatex}{true} 
\usepackage{rotating}
\newboolean{articletitles}
\setboolean{articletitles}{true} 

\newboolean{uprightparticles}
\setboolean{uprightparticles}{false} 
\usepackage{amssymb}
\usepackage{amsfonts}
\usepackage{upgreek}

\usepackage{hyperref}
\usepackage[all]{hypcap} 

\def\lhcb {LHCb\xspace}

\ifthenelse{\boolean{uprightparticles}}%
{

 \def\PDelta      {\ensuremath{\Delta}\xspace}                 
 \def\PXi      {\ensuremath{\Xi}\xspace}                 
 \def\PLambda      {\ensuremath{\Lambda}\xspace}                 
 \def\PSigma      {\ensuremath{\Sigma}\xspace}                 
 \def\POmega      {\ensuremath{\Omega}\xspace}                 
 \def\PUpsilon      {\ensuremath{\Upsilon}\xspace}                 
 
 \def\PB      {\ensuremath{\mathrm{B}}\xspace}                 
                  
 \def\PD      {\ensuremath{\mathrm{D}}\xspace}

 \def\PK      {\ensuremath{\mathrm{K}}\xspace}

 \def\Pc      {\ensuremath{\mathrm{c}}\xspace}

 \def\Pi      {\ensuremath{\mathrm{i}}\xspace}

}
{

 \mathchardef\PDelta="7101
 \mathchardef\PXi="7104
 \mathchardef\PLambda="7103
 \mathchardef\PSigma="7106
 \mathchardef\POmega="710A
 \mathchardef\PUpsilon="7107
                  
 \def\PB      {\ensuremath{B}\xspace}                 
                  
 \def\PD      {\ensuremath{D}\xspace}

 \def\PK      {\ensuremath{K}\xspace}

 \def\Pc      {\ensuremath{c}\xspace}

 \def\Pi      {\ensuremath{i}\xspace}

}

\def\c     {\ensuremath{\Pc}\xspace}

\def\kaon  {\ensuremath{\PK}\xspace}
  \def\Kbar  {\kern 0.2em\overline{\kern -0.2em \PK}{}\xspace}

\def\Kz    {\ensuremath{\kaon^0}\xspace}
\def\Kzb   {\ensuremath{\Kbar^0}\xspace}
\def\KzKzb {\ensuremath{\Kz \kern -0.16em \Kzb}\xspace}
\def\Kp    {\ensuremath{\kaon^+}\xspace}
\def\Km    {\ensuremath{\kaon^-}\xspace}

\def\KpKm  {\ensuremath{\Kp \kern -0.16em \Km}\xspace}

  \def\Dbar    {\kern 0.2em\overline{\kern -0.2em \PD}{}\xspace}
\def\D       {\ensuremath{\PD}\xspace}

\def\Dz      {\ensuremath{\D^0}\xspace}
\def\Dzb     {\ensuremath{\Dbar^0}\xspace}
\def\DzDzb   {\ensuremath{\Dz {\kern -0.16em \Dzb}}\xspace}
\def\Dp      {\ensuremath{\D^+}\xspace}
\def\Dm      {\ensuremath{\D^-}\xspace}

\def\DpDm    {\ensuremath{\Dp {\kern -0.16em \Dm}}\xspace}

  \def\Bbar    {\kern 0.18em\overline{\kern -0.18em \PB}{}\xspace}

  \def\Y#1S{\ensuremath{\PUpsilon{(#1S)}}\xspace}

\newcommand{\tev}{\ensuremath{\mathrm{\,Te\kern -0.1em V}}\xspace}
\newcommand{\gev}{\ensuremath{\mathrm{\,Ge\kern -0.1em V}}\xspace}
\newcommand{\mev}{\ensuremath{\mathrm{\,Me\kern -0.1em V}}\xspace}
\newcommand{\kev}{\ensuremath{\mathrm{\,ke\kern -0.1em V}}\xspace}
\newcommand{\ev}{\ensuremath{\mathrm{\,e\kern -0.1em V}}\xspace}
\newcommand{\gevc}{\ensuremath{{\mathrm{\,Ge\kern -0.1em V\!/}c}}\xspace}
\newcommand{\mevc}{\ensuremath{{\mathrm{\,Me\kern -0.1em V\!/}c}}\xspace}
\newcommand{\gevcc}{\ensuremath{{\mathrm{\,Ge\kern -0.1em V\!/}c^2}}\xspace}
\newcommand{\gevgevcccc}{\ensuremath{{\mathrm{\,Ge\kern -0.1em V^2\!/}c^4}}\xspace}
\newcommand{\mevcc}{\ensuremath{{\mathrm{\,Me\kern -0.1em V\!/}c^2}}\xspace}

\def\to                 {\ensuremath{\rightarrow}\xspace}

\def\gsim{{~\raise.15em\hbox{$>$}\kern-.85em
          \lower.35em\hbox{$\sim$}~}\xspace}
\def\lsim{{~\raise.15em\hbox{$<$}\kern-.85em
          \lower.35em\hbox{$\sim$}~}\xspace}

\def\pythia     {\mbox{\textsc{Pythia}}\xspace}

\def\AT#1     {\ensuremath{A_T^{#1}}\xspace}           

\def\C#1      {\ensuremath{\mathcal{C}_{#1}}}                       
\def\Cp#1     {\ensuremath{\mathcal{C}_{#1}^{'}}}                    
\def\Ceff#1   {\ensuremath{\mathcal{C}_{#1}^{\mathrm{(eff)}}}}        
\def\Cpeff#1  {\ensuremath{\mathcal{C}_{#1}^{'\mathrm{(eff)}}}}       
\def\Ope#1    {\ensuremath{\mathcal{O}_{#1}}}                       
\def\Opep#1   {\ensuremath{\mathcal{O}_{#1}^{'}}}                    

\usepackage{bm}
\usepackage{afterpage}
\usepackage{mciteplus}
\begin{document}
\renewcommand{\thefootnote}{\fnsymbol{footnote}}
\setcounter{footnote}{1}



\begin{titlepage}

\belowpdfbookmark{Title page}{title}

\pagenumbering{roman}
\vspace*{-1.5cm}
\centerline{\large EUROPEAN ORGANIZATION FOR NUCLEAR RESEARCH (CERN)}
\vspace*{1.5cm}
\hspace*{-5mm}\begin{tabular*}{16cm}{lc@{\extracolsep{\fill}}r}
\vspace*{-12mm}\mbox{\!\!\!\includegraphics[width=.12\textwidth]{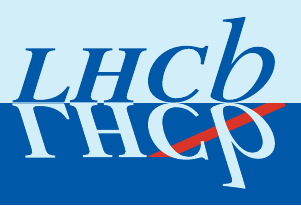}}& & \\
 & & CERN-PH-EP-2012-006\\
 & & LHCb-PAPER-2011-038 \\  
 & & January 26, 2012\\ 
 & & \\
\hline
\end{tabular*}

\vspace*{3.0cm}

{\bf\boldmath\Large
\begin{center}
  Searches for Majorana neutrinos in $B^-$ decays\\
\end{center}
}

\vspace*{2cm}
\begin{center}
\normalsize {
The LHCb Collaboration\footnote{Authors are listed on the following pages.}
}
\end{center}


\begin{abstract}
  \noindent  
  
Searches for heavy Majorana neutrinos in $B^-$ decays in final states containing hadrons plus a $\mu^-\mu^-$ pair have been performed using 0.41 fb$^{-1}$ of data collected with the LHCb detector in proton-proton collisions at a center-of-mass energy of 7~TeV.  
The $D^{+} \mu^- \mu^-$  and $D^{*+} \mu^- \mu^-$  final states can arise from the presence of virtual Majorana neutrinos of any mass. Other final states containing  $\pi^+$, $D_s^+$, or $D^0\pi^+$ can be mediated by an on-shell Majorana neutrino.
 No signals are found and upper limits are set on Majorana neutrino production as a function of mass, and also on the $B^-$ decay branching fractions. 

\end{abstract}

\vspace*{2.0cm}
\vspace{\fill}

\vspace*{1.0cm}
\hspace*{6mm}{\it PACS:} 14.40.Nd, 13.35.Hb, 14.60.Pq \\
\hspace*{6mm}Submitted to Physics Review D\\
\newpage
\begin{center}
The LHCb Collaboration\\
\begin{flushleft}
\small
R.~Aaij$^{38}$, 
C.~Abellan~Beteta$^{33,n}$, 
B.~Adeva$^{34}$, 
M.~Adinolfi$^{43}$, 
C.~Adrover$^{6}$, 
A.~Affolder$^{49}$, 
Z.~Ajaltouni$^{5}$, 
J.~Albrecht$^{35}$, 
F.~Alessio$^{35}$, 
M.~Alexander$^{48}$
% ...

{\footnotesize \it
$ ^{1}$Centro Brasileiro de Pesquisas F\'{i}sicas (CBPF), Rio de Janeiro, Brazil\\
% ...
$ ^{5}$Clermont Universit\'{e}, Universit\'{e} Blaise Pascal, CNRS/IN2P3, LPC, Clermont-Ferrand, France\\
$ ^{6}$CPPM, Aix-Marseille Universit\'{e}, CNRS/IN2P3, Marseille, France\\
% ...
$ ^{33}$Universitat de Barcelona, Barcelona, Spain\\
$ ^{34}$Universidad de Santiago de Compostela, Santiago de Compostela, Spain\\
$ ^{35}$European Organization for Nuclear Research (CERN), Geneva, Switzerland\\
% ...
$ ^{38}$Nikhef National Institute for Subatomic Physics, Amsterdam, The Netherlands\\
% ...
$ ^{43}$H.H. Wills Physics Laboratory, University of Bristol, Bristol, United Kingdom\\
% ...
$ ^{48}$School of Physics and Astronomy, University of Glasgow, Glasgow, United Kingdom\\
$ ^{49}$Oliver Lodge Laboratory, University of Liverpool, Liverpool, United Kingdom\\
% ...
$ ^{n}$LIFAELS, La Salle, Universitat Ramon Llull, Barcelona, Spain\\
% ...
}
\end{flushleft}

\end{center}

\end{titlepage}

\renewcommand{\thefootnote}{\arabic{footnote}}
\setcounter{footnote}{0}
\pagestyle{empty}  



%





\pagestyle{plain} 
\setcounter{page}{1}
\pagenumbering{arabic}


%


\section{Introduction}
\label{sec:Introduction}

\begin{figure}[hbt]
\centering
\includegraphics[width=4.in]{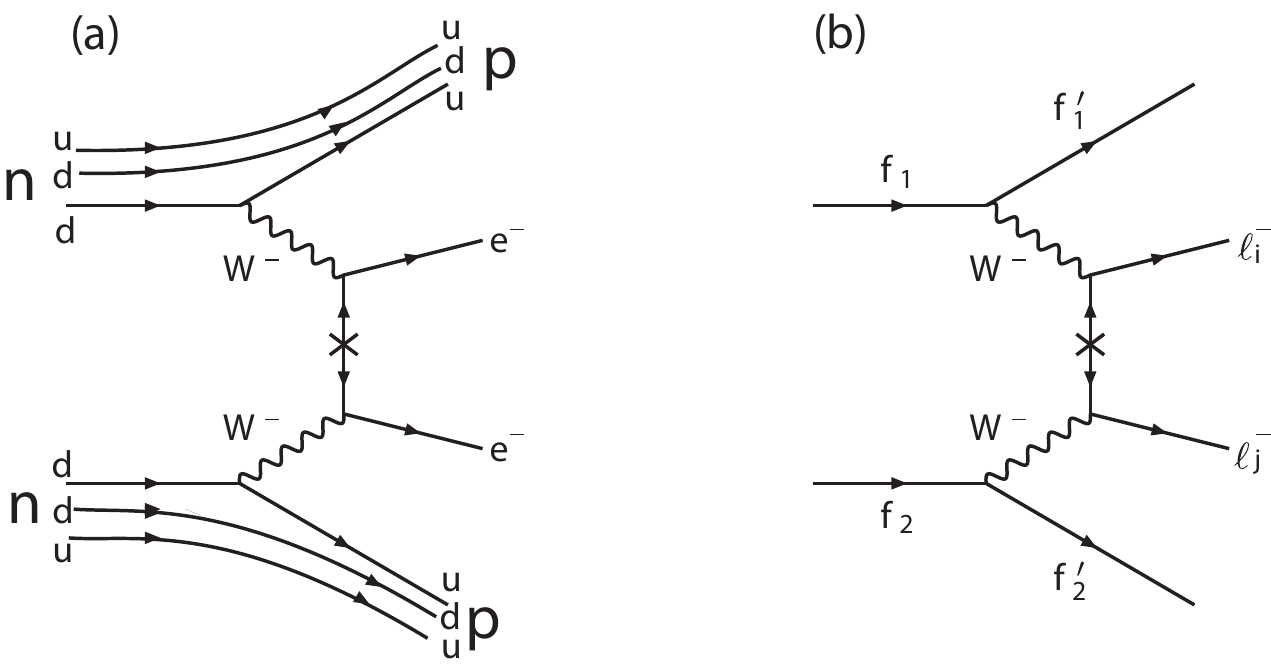}
\caption{(a) Diagram of neutrinoless double $\beta$ decay when two neutrons in a nucleus decay simultaneously. (b) The fundamental diagram for changing lepton number by two units.
} \label{doublebeta}
\end{figure}

Leptons constitute a crucially important sector of elementary particles. Half of the leptons are neutrinos. Yet we do not know if they
are Dirac or Majorana particles, the latter case characterized by being their own antiparticles \cite{Majorana}. 
Since the observation of neutrino oscillations has indisputably established that neutrinos have non-zero mass, it is possible to distinguish the two types experimentally.
Finding neutrinoless double $\beta$ decay has long been advocated as a premier demonstration of the possible  Majorana nature of neutrinos \cite{Avignone:2007fu}. 
The Feynman diagram is shown in Fig.~\ref{doublebeta}. We also show the fundamental quark and lepton level process. An impressive lower limit from neutrinoless double $\beta$ decays in nuclei has already been obtained on the half-life of $\cal{O}$$(10^{25})$ years~\cite{PDG} for coupling to $e^-$. 

\begin{figure}[hbt]
\centering
\includegraphics[width=5.in]{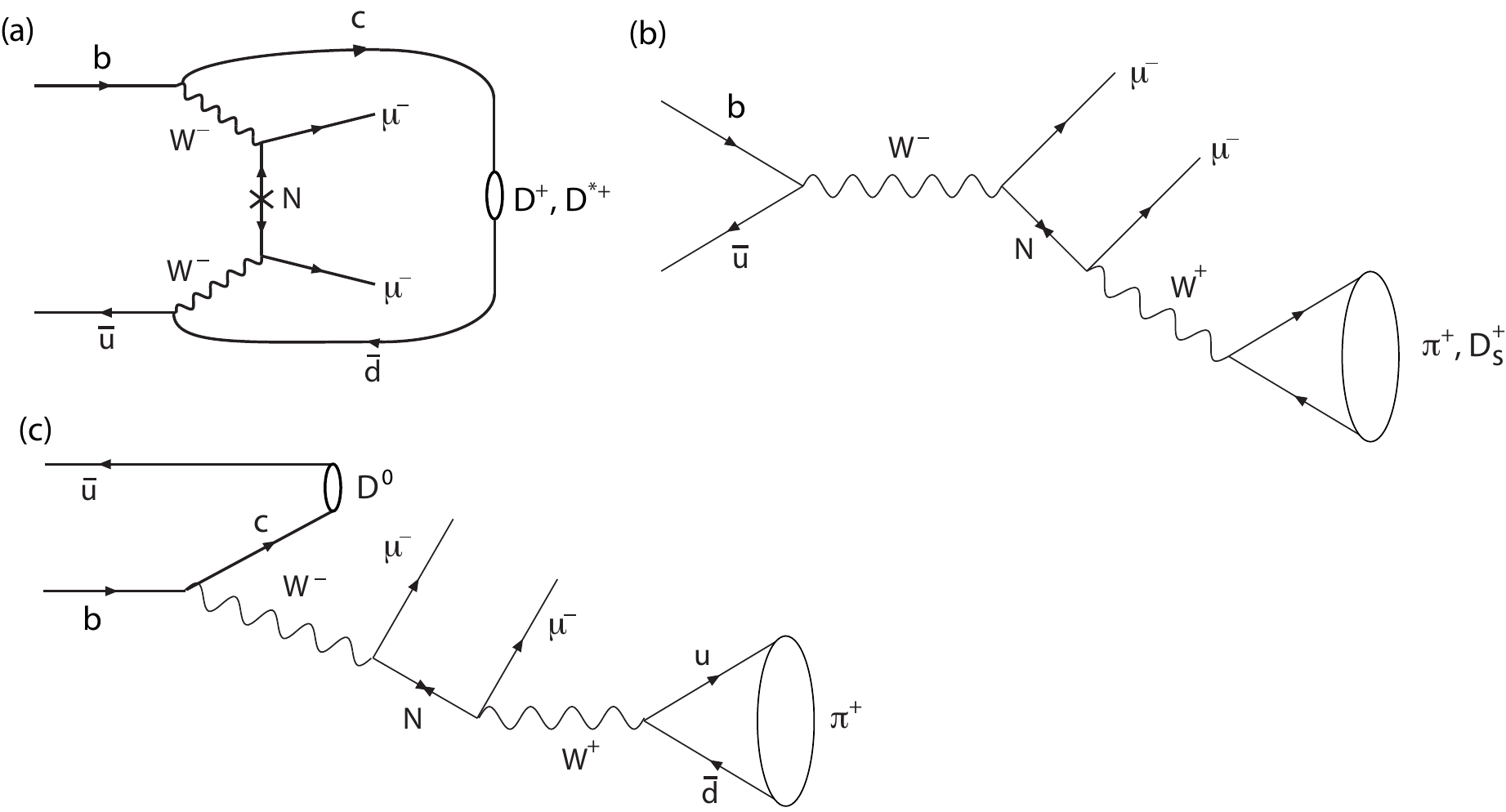}
\caption{Feynman diagrams for $B$ decays involving an intermediate heavy neutrino ($N$). (a)  $B^-\to D^{(*)+}\mu^-\mu^-$, (b) $B^-\to \pi^+(D_s^+)\mu^-\mu^-$,  and (c) $B^-\to D^0 \pi^+\mu^-\mu^-$.  } \label{B-Majorana}
\end{figure}

Similar processes can occur in $B^-$ decays. The diagram is shown in Fig.~\ref{B-Majorana}(a). In this reaction there is no restriction on the mass of the Majorana neutrino as it acts as a virtual particle. In this paper,
unlike in neutrinoless double beta decays, a like-sign dimuon is considered rather than two electrons.
The only existing limit is from a recent Belle measurement \cite{Seon:2011ni} using the $B^- \to D^+\mu^- \mu^- $ channel.
We consider only final states where the $c\overline{d}$ pair forms a final-state meson, either a $D^+$ or a $D^{*+}$, so the processes we are looking for are $B^-\to D^{(*)+} \mu^-\mu^-$. 
In this paper mention of a specific reaction also implies inclusion of the charge conjugate reaction. 

There are other processes involving $b$-quark decays that produce a light neutrino that can mix with a heavy neutrino, designated as $N$.  The heavy neutrino can  decay as $N\to W^+\mu^-$. In  Fig.~\ref{B-Majorana}(b) we show the annihilation processes $B^-\to \pi^+(D_s^+)\mu^-\mu^- $, where the virtual $W^+$ materializes either as a $\pi^+$ or $D_s^+$. These decays have been discussed in the literature \cite{Atre:2009rg,Cvetic:2010rw,*Zhang:2010um}.

We note that it is also possible for the $B^-\to D^{(*)+} \mu^-\mu^-$ decay modes shown in Fig.~\ref{B-Majorana}(a)
to proceed by a Cabibbo suppressed version of the process in Fig.~\ref{B-Majorana}(b) where the virtual $W^+$ forms $D^{(*)+}$.
Similarly, the decay modes shown in Fig.~\ref{B-Majorana}(b) could be produced
via Cabibbo suppressed versions of the process in Fig.~\ref{B-Majorana}(a).  Here the $\pi^+\mu^-\mu^-$ final state requires a $b\to u$ quark transition while for the $D_s^+\mu^-\mu^-$ final state, one of the virtual $W^-$ must couple to a $\overline{s}$ quark rather than a $\overline{d}$.

The lifetimes of $N$ are not predicted. We assume here that they are long enough that the natural decay width is narrower than our mass resolution which varies between 2 and 15 MeV\footnote{In this paper we use units where the speed of light, $c$, is set equal to one.}  depending on mass and decay mode.
For $B^-\to \pi^+ \mu^-\mu^- $, we can access the Majorana mass region between approximately 260 and 5000 MeV while for $B^-\to D_s^+\mu^-\mu^- $,  the Majorana mass region is between 2100 and 5150 MeV.
In the higher mass region,  the $W^+$ may be more likely to form a $D_s^+$ meson than a $\pi^+$.
The $B^- \to \pi^+\mu^- \mu^-$ search was first performed by Mark-\uppercase\expandafter{\romannumeral2}~\cite{Weir:1989sq} and then by CLEO~\cite{Edwards:2002kq}.
LHCb also performed a similar search using a smaller 0.04~fb$^{-1}$ data sample \cite{Aaij:2011ex} giving an upper limit of $5.8 \times 10^{-8}$ at 95\% confidence level (CL).
The decay of $B^-\to D_s^+\mu^-\mu^- $ has never been investigated.

Finally, in Fig.~\ref{B-Majorana}(c) we show how prolific semileptonic decays of the $B^-$ can result in the $D^0\pi^+\mu^-\mu^-$ final state. 
This process has never been probed \cite{Quintero:2011yh}.
We benefit from the higher value of the CKM coupling $|V_{cb}|$ relative to  $|V_{ub}|$ in the annihilation processes shown in  Fig.~\ref{B-Majorana}(b).
The accessible region for Majorana neutrino mass is between 260 and 3300 MeV. For all the modes considered in this paper, we search only for decays with muons in the final state, though electrons, and $\tau$ leptons in cases where sufficient energy is available, could also be produced.
Searches have also been carried out looking for like-sign dileptons in hadron collider experiments \cite{Aad:2011vj,*Chatrchyan:2011wba,*Abulencia:2007rd,*Acosta:2004kx}.

\section{Data sample and signal}
We use a data sample of 0.37\,fb$^{-1}$ collected with the LHCb detector  \cite{LHCb-det} in the first half of 2011 and an additional
0.04\,fb$^{-1}$ collected in 2010 at a center-of-mass energy of 7 TeV. 

The detector elements are placed along the beam line of the LHC starting with the vertex detector, a silicon strip device that surrounds the proton-proton interaction region having its first active layer positioned 8 mm from the beam during collisions. It provides precise locations for primary $pp$ interaction vertices, the locations of decays of long-lived particles, and contributes to the measurement of track momenta. Further downstream, other devices used to measure track momenta include a large area silicon strip detector located in front of a 4 Tm dipole magnet, and a combination of
silicon strip detectors and straw-tube drift chambers placed behind. Two Ring Imaging Cherenkov (RICH) detectors are used to identify charged hadrons. An electromagnetic calorimeter is used for photon detection and electron identification, followed by a hadron calorimeter, and  a system that distinguishes muons from hadrons. The calorimeters and the muon system provide first-level hardware triggering, which is then followed by a software high level trigger.

Muons are triggered on at the hardware level using their penetration through iron and detection in a series of tracking chambers. Projecting these tracks through the magnet to the primary event vertex allows a determination of their  transverse momentum, $p_{\rm T}$. Events from the 2011 data used in this analysis were triggered on the basis of a single muon having a $p_{\rm T}$ greater than 1480\,MeV, or two muons with their product $p_{\rm T}$ greater than 1.69\,GeV$^{2}$.  To satisfy the higher level trigger, the muon candidates must also be detached from the primary vertex. 

Candidate $B^-$ decays
are found using tracking information, and particle identification information from the RICH and muon systems. 
The identification of pions, kaons and muons is based on combining the information from the two RICH detectors, the calorimeters and the muon system. The RICH detectors measure the angles of emitted Cherenkov radiation with respect to each charged track. For a given momentum particle this angle is known, so a likelihood for each hypothesis is computed. Muon likelihoods are computed based on track hits in each of the sequential muon chambers. In this analysis we do not reject candidates based on sharing hits with other tracks. This eliminates a possible bias that was present in our previous analysis \cite{Aaij:2011ex}.
Selection criteria are applied on the difference of the logarithm of the likelihood between two hypotheses. The 
efficiencies and the mis-identification rates are obtained from data using $K_S$, $D^{*+}\to \pi^+ D^0$, $D^0\to K^-\pi^+$ and $J/\psi\to\mu^+\mu^-$ event samples that provide almost pure pion, kaon, and muon sources.

Efficiencies and and rejection rates depend on the momentum of the final state particles. For the RICH detector generally the pion or kaon efficiencies exceed 90\% and the rejection rates are of the order of 5\% \cite{Powell:2011zz}. The muon system provides efficiencies exceeding 98\% with rejection rates on hadrons of better than 99\%, depending on selection criteria \cite{LHCb-PUB-2011-027}.
 Tracks of good quality are selected for further analysis. In order to ensure that tracks have good vertex resolution we insist that they all have $p_{\rm T}>300$\,MeV. For muons this requirement varies from 650$-$800\,MeV depending on the final state. All tracks must be inconsistent with having been 
produced at the primary vertex closest to the candidate $B^-$ meson's decay point. The impact parameter (IP) is the minimum distance of approach of the track with respect to the primary vertex. Thus we form the IP $\chi^2$  by testing the hypothesis that the IP is equal to zero, and require it to be large; the values depend on the decay mode and range from 4 to 35.


\begin{table}
\begin{center}
\caption{Charm and charmonium branching fractions}\label{tab:charms}
\begin{tabular}{ccrl}\hline\hline
Particle & Final state &\multicolumn{2}{c}{Branching fraction (\%)}\\\hline
$D^0$ & $K^-\pi^+$  &  ~~~~3.89$\pm$&\!\!\!\!\!\!0.05 \cite{PDG} \\
$D^+$ & $K^-\pi^+\pi^+$ & 9.14$\pm$&\!\!\!\!\!\!0.20 \cite{PDG} \\
$D_s^+$ & $K^-K^+\pi^+$  &5.50$\pm$&\!\!\!\!\!\!0.27 \cite{:2008cqa} \\
$D^{*+}$ & $\pi^+D^0$  &67.7$\pm$&\!\!\!\!\!\!0.5 \cite{PDG} \\
$\psi(2S)$ & $\pi^+\pi^-J/\psi$ & 32.6$\pm$&\!\!\!\!\!\!0.5 \cite{PDG} \\
$J/\psi$ &$\mu^+\mu^-$ & 5.93$\pm$&\!\!\!\!\!\!0.06 \cite{PDG} \\
\hline\hline
\end{tabular}
\end{center}
\end{table}

\section{Normalization channels}
\label{sec:norm}
Values for branching fractions will be normalized to well measured channels that have the same number of muons in the final state and equal track multiplicities. The first such channel is $B^- \to J/\psi K^-$. Its  branching fraction is ${\cal{B}}(B^- \to J/\psi K^-)= (1.014 \pm 0.034) \times 10^{-3}$ \cite{PDG}. We use the $J/\psi\to \mu^+\mu^-$ decay mode. The product branching fraction of this normalization channel is $(6.013\pm 0.021)\times 10^{-5}$, and is known to an accuracy of $\pm$2\%.
 The charm meson decay modes used in this paper are listed in Table~\ref{tab:charms}, along with their branching fractions and those of the charmonium decays in the normalization channels. 
 
To select the  $J/\psi K^-$ normalization channel,
the $p_{\rm T}$ requirement is increased to 1100\,MeV for the $K^-$ and 750\,MeV for the muons.
To select $B^-$ candidates we further require that the three tracks form a vertex with a $\chi^2< 7$, and that this $B^-$ candidate points to the primary vertex at an angle not different from its momentum direction by more than 4.47\,mrad, and that the impact parameter $\chi^2$ of the $B^-$ is less than 12. The same requirements will be used for the $\pi^+\mu^-\mu^-$ selection. The total efficiency for $\mu^+\mu^-K^-$ is (0.99$\pm$0.01)\%, where the $\mu^+\mu^-$ come from $J/\psi$ decay.

\begin{figure}[hbt]
\centering
\includegraphics[width=6in]{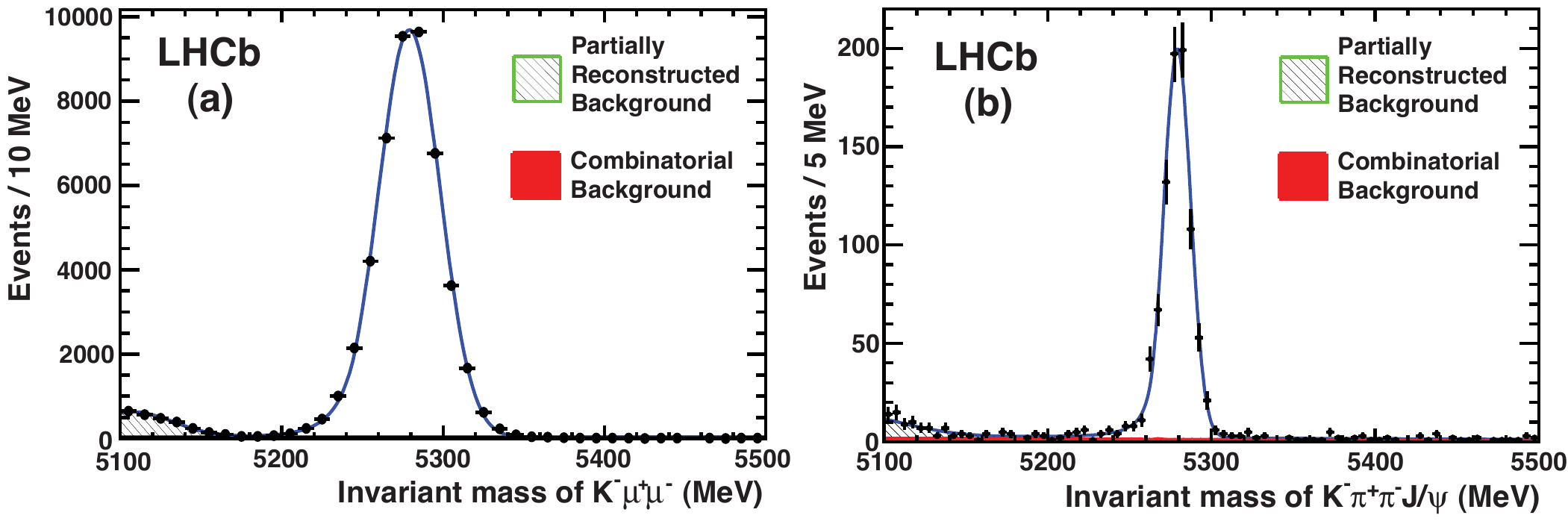}
\caption{Invariant mass of (a) candidate $J/\psi K^-$ decays, and (b) candidate $J/\psi K^-\pi^+\pi^-$ decays. The data are shown as the points with error bars. Both the partially reconstructed background 
and the combinatorial background are shown, although the combinatorial background is small and barely visible. The solid curve shows the total. In both cases the candidate $\mu^+\mu^-$ is required to be within $\pm$50 MeV of the $J/\psi$ mass, and in (b) the dimuon pair is constrained to have the $J/\psi$ mass.}
\label{fig:JpsiK}
\end{figure}

The invariant mass of $K^-\mu^+\mu^-$ candidates is shown in Fig.~\ref{fig:JpsiK}(a). In this analysis the  $\mu^+\mu^-$ invariant mass is required to be within 50 MeV of the $J/\psi$ mass. 
We use a Crystal Ball function (CB) to describe the signal \cite{Skwarnicki:1986xj}, a Gaussian distribution for the partially reconstructed background events,  and a linear distribution for combinatorial background. The CB function provides a convenient way to describe the shape of the distribution, especially in the mass region below the peak where radiative effects often produce an excess of events that falls away gradually, a so called ``radiative tail".
The CB function is
\begin{equation}
f(m;\alpha,n, {m}_0,\sigma) = \left\{
\begin{array}{l l}
 \exp\left(-\frac{(m-m_0)^2}{2\sigma^2}\right)  & \quad \mbox{for $\frac{m-m_0}{\sigma}>-\alpha$}\\
  A\cdot \left(b-\frac{m-m_0}{\sigma}\right)^{-n} & \quad \mbox{for $\frac{m-m_0}{\sigma}\le-\alpha$}\\ \end{array} \right. 
  \end{equation}
 where
 \[A=\left(\frac{n}{|\alpha|}\right)^n\cdot\exp\left(-\frac{|\alpha|^2}{2}\right)\]
 \begin{equation}
b=\frac{n}{|\alpha|}-|\alpha|.\nonumber
\end{equation}
The measured mass of each candidate is indicated as $m$, while $m_0$ and  $\sigma$ are the fitted peak value and resolution, and $n$ and $\alpha$ are parameters used to model the radiative tail.  We use the notation $\sigma$ in the rest of this paper to denote resolution values found from CB fits.

Using an unbinned log-likelihood fit yields 47,224$\pm$222 $B^- \to J/\psi K^-$ events. Within a $\pm 2\sigma$ signal window about the peak mass, taken as the signal region, there are 44,283 of these events.  The number of signal events in this window is also determined using the total number of events and subtracting the number given by the background fit. 
The difference is 119 events, and this is taken as the systematic uncertainty of 0.3\%.
 The width of the signal peak is found to be 19.1$\pm$0.1 MeV. Monte Carlo simulations are based on event generation using \pythia  \cite{Sjostrand:2006za}, followed by a G{\sc eant-4} \cite{Agostinelli:2002hh} based simulation of the LHCb detector \cite{LHCb-PROC-2011-006}. The $J/\psi K^-$ mass resolution is
 20\% larger than that given by the LHCb simulation.  All simulated mass resolutions in this paper are increased by this factor.

For final states with five tracks, we change the normalization channel to $B^-\to \psi(2S)K^-$, with $\psi(2S)\to \pi^+\pi^-J/\psi$, and $J/\psi\to \mu^+\mu^-$. The branching fraction for this channel is ${\cal{B}}(B^-\to \psi(2S)K^-)=(6.48\pm 0.35)\times 10^{-4}$ \cite{PDG}.
Events are selected using a similar procedure as for $J/\psi K^-$ but adding a $\pi^+\pi^-$ pair, that must have an invariant mass when combined with the $J/\psi$ which is compatible with the $\psi(2S)$ mass, and that forms a consistent vertex with the other $B^-$ decay candidate tracks. The total efficiency for $\mu^+\mu^-\pi^+\pi^- K^-$ is (0.078$\pm$0.002)\%, without inclusion of the $\psi(2S)$ or $J/\psi$ branching fractions.
 The $B^-$ candidate mass plot is shown in Fig.~\ref{fig:JpsiK}(b). Here the $\mu^+\mu^-$ pair is constrained to the $J/\psi$ mass. (In what follows, whenever the final state contains a ground-state charm meson, its decay products are constrained to their respective charm masses.) 

The data are fitted with a CB function for signal, a Gaussian distribution for partially reconstructed background and a linear function for combinatorial background. There are
767$\pm$29 signal events in a $\pm 2\sigma$ window about the peak mass. The difference between this value and a count of the number of events in the signal region after subtracting the background implies a 0.7\% systematic uncertainty on the yield. 

\section{\boldmath Analysis of $B^-\to D^+\mu^-\mu^-$ and $D^{*+}\mu^-\mu^-$}
Decay diagrams for $B^-\to D^{(*)+}\mu^-\mu^-$ are shown in Fig.~\ref{B-Majorana}(a).  Since the neutrinos are virtual, the process can proceed for any value of neutrino mass. It is also possible for these decays to occur via a Cabibbo suppressed process similar to the ones shown in Fig.~\ref{B-Majorana}(b), where the virtual $W^+$ materializes as a $c\overline{d}$ pair. If this occurred we would expect the Cabibbo allowed $D_s^+\mu^-\mu^-$ final state to be about an order of magnitude larger. The search for Majorana neutrinos in this channel are discussed in Section~\ref{sec:Dsmumu}. The $D^+\to K^-\pi^+\pi^+$ and $D^{*+}\to \pi^+ D^0,~D^0\to K^-\pi^+$ channels are used. The decay products of the $D^+$ and $D^0$ candidates are required to have invariant masses within $\pm$25 MeV of the charm meson mass, and for $D^{*+}$ candidate selection the mass difference  $m(\pi^+ K^-\pi^+)-m(K^-\pi^+)$ is required to be within $\pm3$ MeV of the known $D^{*+}-D^0$ mass difference.

\begin{figure}[hbt]
\centering
\includegraphics[width=6in]{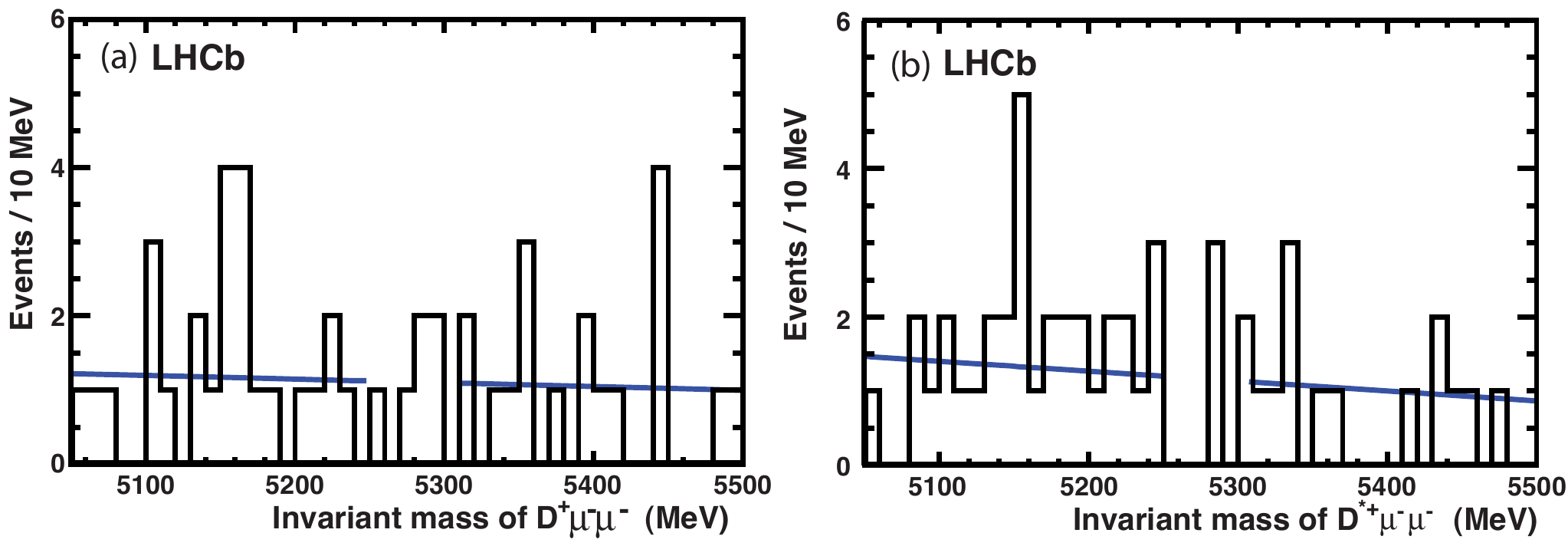} 
\caption{Invariant mass spectrum for (a) $B^-\to D^+\mu^-\mu^-$ candidates, and (b) $B^-\to D^{*+}\mu^-\mu^-$ candidates. The solid lines show the linear fits to the data in the mass sidebands. } \label{fig: MassFitDp}
\end{figure}

The $D^{(*)+}\mu^-\mu^-$ candidate mass spectra are shown in Fig.~\ref{fig: MassFitDp}. No signals are apparent.  The $B^-$ mass resolution is 15.7$\pm$0.5 MeV for the $D^+$ channel and 14.1$\pm$0.6 MeV for the $D^{*+}$ channel. 
The background has two components, one from mis-reconstructed $B$ decays that tends to peak close to the $B^-$ mass, called ``peaking backgrounds", and random track combinations that are parameterized by a linear function. 
To predict the combinatorial background in the signal region we fit the data in the sidebands with a straight line. In the $D^+$ mode we observe six events in the signal region, while there are five in the $D^{*+}$ mode.
The combinatorial background estimates are 6.9$\pm$1.1 and 5.9$\pm$1.0 events, respectively.  Peaking backgrounds are estimated from mis-identification probabilities, determined from data, coupled with Monte Carlo simulation. For these two channels peaking backgrounds are very small. The largest, due to $B^-\to D^+\pi^-\pi^-$, is only 0.04 events.

The total efficiencies for $D^+\mu^-\mu^-$ and $D^{*+}\mu^-\mu^-$ are $(0.099\pm 0.007)$\% and $(0.066\pm 0.005)$\%, respectively; here the charm branching fractions are not included.
The systematic errors are listed in Table~\ref{tab: SysD} for this mode and other modes containing charm mesons that will be discussed subsequently. Trigger efficiency uncertainties are evaluated from differences in the 2010 and 2011 data samples. The largest systematic uncertainties are due to the branching fractions of the normalization channels and the trigger efficiencies. 
The uncertainty on the background is taken into account directly when calculating the upper limits as explained below. Other uncertainties arise from errors on the charmed meson branching fractions. For these final states the uncertainty due to different final state track momenta with respect to the normalization mode are very small, on the order of 0.2\%.
Other channels have uncertainties due to varying efficiencies as a function of Majorana mass, and these are entered in the row labeled ``Efficiency modeling". The detector efficiency modeling takes into account the different acceptances that could be caused by having different track momentum spectra. For example, the track momenta depend on the Majorana neutrino mass for on-shell neutrinos. 
These uncertainties are ascertained by simulating the detector response at fixed Majorana masses and finding the average excursion from a simple fit to the response and the individually simulated mass points. This same method is used for other modes.

\begin{table}[!htb]
\centering
\caption{Systematic uncertainties for $B^- \to D X \mu^- \mu^-$ modes. }
\label{tab: SysD}
\begin{tabular}{lcccc}
\hline\hline
Source & \multicolumn{4}{c}{Systematic uncertainty (\%)}\\\hline
Common to all modes & \multicolumn{4}{c}{} \\\hline
${\cal{B}}(B^- \to \psi(2S) K^-)$ &\multicolumn{4}{c}{5.4}\\
${\cal{B}}(\psi(2S) \to J/\psi \pi^+ \pi^-)$ & \multicolumn{4}{c}{1.5}\\
${\cal{B}}(J/\psi \to \mu^+ \mu^-)$ &\multicolumn{4}{c}{1.0}\\
Uncertainty in signal shape &\multicolumn{4}{c}{3.0}\\
Yield of reference channel & \multicolumn{4}{c}{0.7}\\
$\mu$ PID  &\multicolumn{4}{c}{0.6} \\\hline
Mode specific             & $D_s^+$ & $D^0\pi^+$& $D^+$ & $D^{*+}$\\\hline
Trigger& 4.9   & 9.3 & 5.5 &4.8 \\
Efficiency modeling& 10.0 & 6.7 && \\
PID ($K/\pi$)& 1.0&&& \\
Charm decay ${\cal{B}}$'s & 4.9 & 1.3 & 2.2 & 1.5\\\hline
Total &13.8 & 13.2&8.8 &8.2\\\hline\hline
\end{tabular}
\end{table}

To set upper limits on the branching fraction the number of events $N_{\rm obs}$ within $\pm2\sigma$ of the $B^-$ mass are counted. The distributions of the number of events ($N$) are Poisson with the mean value of ($S+B$), where $S$ indicates the expectation value of signal and $B$ background.
For a given number of observed events in the signal region, the upper limit is calculated using the probability for $N \le N_{\rm obs}$:
\begin{equation}
P(N \leq N_{\rm obs}) =\sum_{N \le N_{\rm obs}} \frac{(S+B)^N e^{-(S+B)}}{N!}.
\end{equation}
A limit at 95\% CL for branching fraction calculations is set by having 
$P(N \leq N_{\rm obs}) = 0.05.$ The systematic errors are taken into account by 
varying the
calculated $S$ and $B$, assuming Gaussian distributions.

The upper limits on the branching fractions at 95\% CL are measured to be
\begin{eqnarray}
{\cal{B}}(B^-\to D^+ \mu^-\mu^-)&<& 6.9\times 10^{-7}~{\rm and} \nonumber\\
{\cal{B}}(B^-\to D^{*+}\mu^-\mu^-)&<& 2.4\times 10^{-6}~. \nonumber
\end{eqnarray}
The limit on the $D^+$ channel is more stringent than a previous limit from Belle  of $1 \times 10^{-6}$ at 90\% CL \cite{Seon:2011ni}, and the limit on the $D^{*+}$ channel is the first such result.

\section{\boldmath Analysis of $B^-\to \pi^+\mu^-\mu^-$}

\begin{figure}[hbt]
\centering
\includegraphics[width=4.5 in]{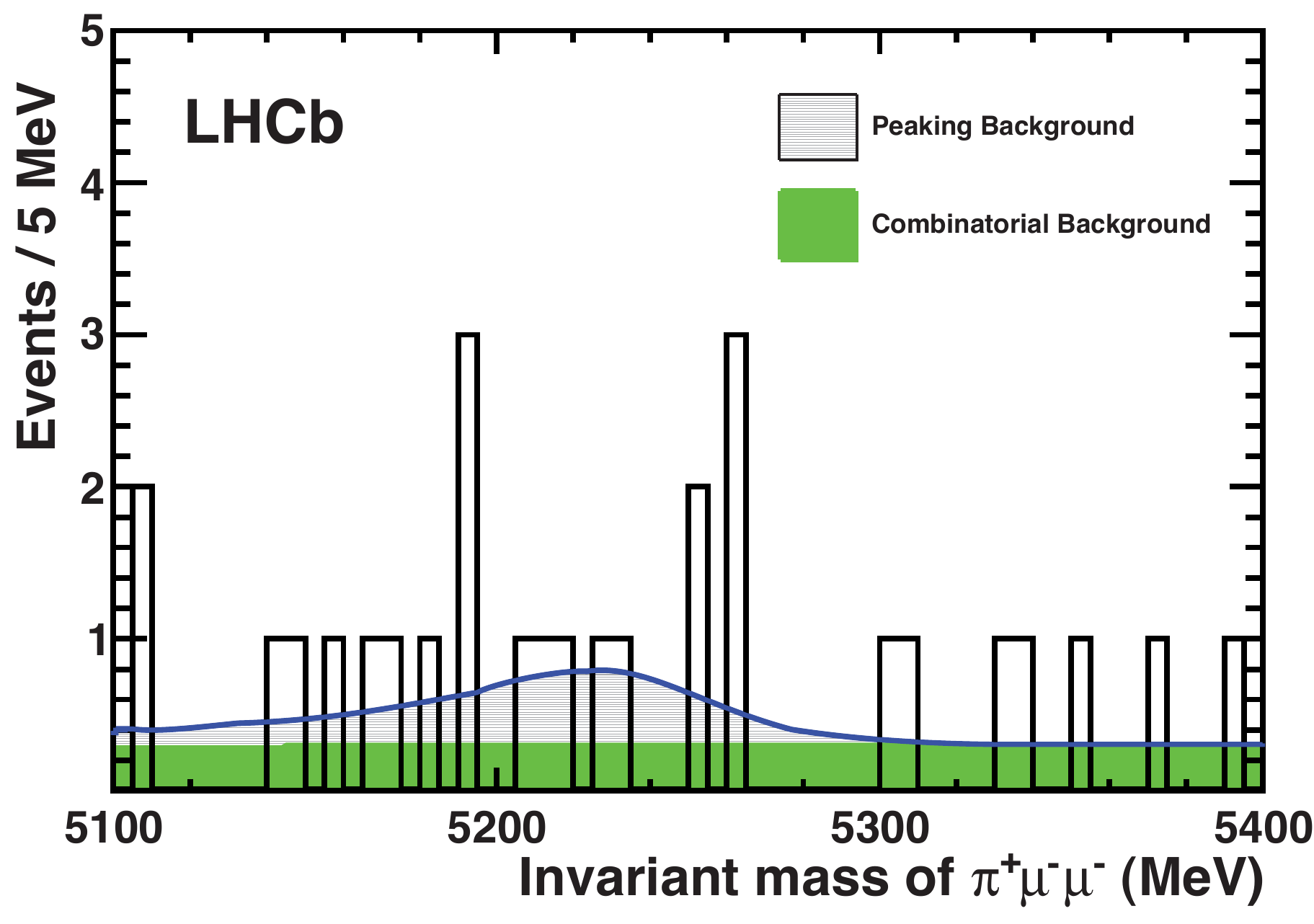}
\caption{Invariant mass distribution of $\pi^+\mu^-\mu^-$. The estimated backgrounds are also shown. The curve
is the sum of the peaking background and the combinatoric background. } \label{MassFitPi}
\end{figure}

The selection of $\pi^+\mu^-\mu^-$ events uses the same criteria as described for $J/\psi K^-$ in Section~\ref{sec:norm}, except for like-sign rather than opposite-sign dimuon charges and pion rather than kaon identification.
The invariant mass distribution of $\pi^+\mu^-\mu^-$ candidates is shown in Fig.~\ref{MassFitPi}. The mass resolution for this final state is 20.3$\pm$0.2 MeV.  An interval of $\pm$2$\sigma$ centered on the $B^-$ mass is taken as the signal region. 
There are 7 events in the signal region, but no signal above background is apparent. The peaking background, estimated as 2.5 events, 
is due to misidentified $B^-\to J/\psi K^-$  or $J/\psi \pi^-$ decays;  the shape is taken from simulation.  The combinatorial background is determined to be 5.3 events from a fit to the $\pi^+\mu^-\mu^-$ mass distribution excluding the signal region. The total background in the signal region then is $7.8\pm1.3$ events.

\begin{figure}[hbt]
\centering
\includegraphics[width=4.5 in]{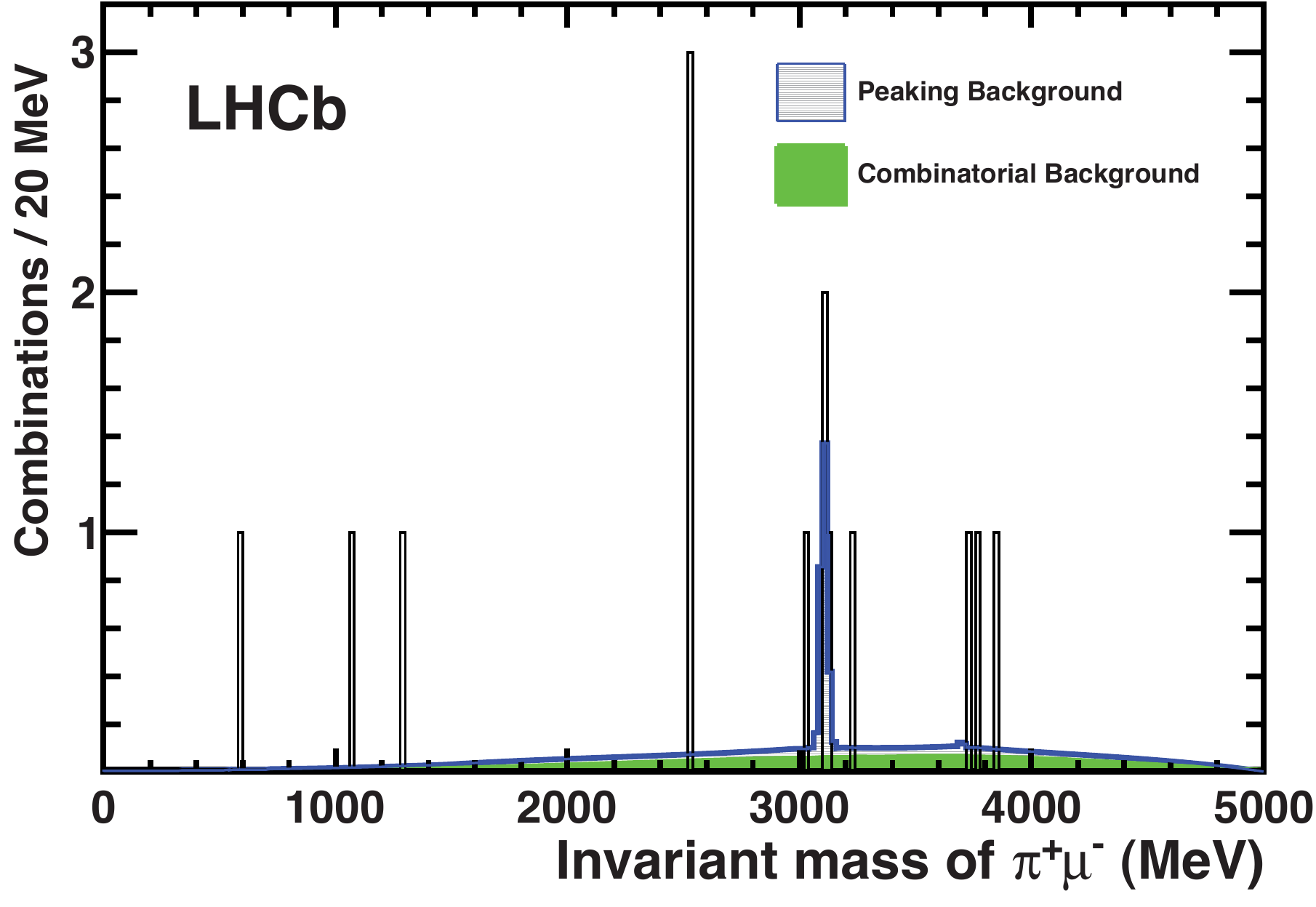}
\caption{Invariant mass distribution of $\pi^+\mu^-$ in the $\pm 2\sigma$ region of the $B^-$ mass with both peaking and combinatorial background superimposed. The peaking background at 3100 MeV is due to misidentified $B^-\to J/\psi X$ decays.There are two combinations per event.} \label{MassSignalN}
\end{figure}

\begin{figure}[hbt]
\centering
\includegraphics[width=4.5in]{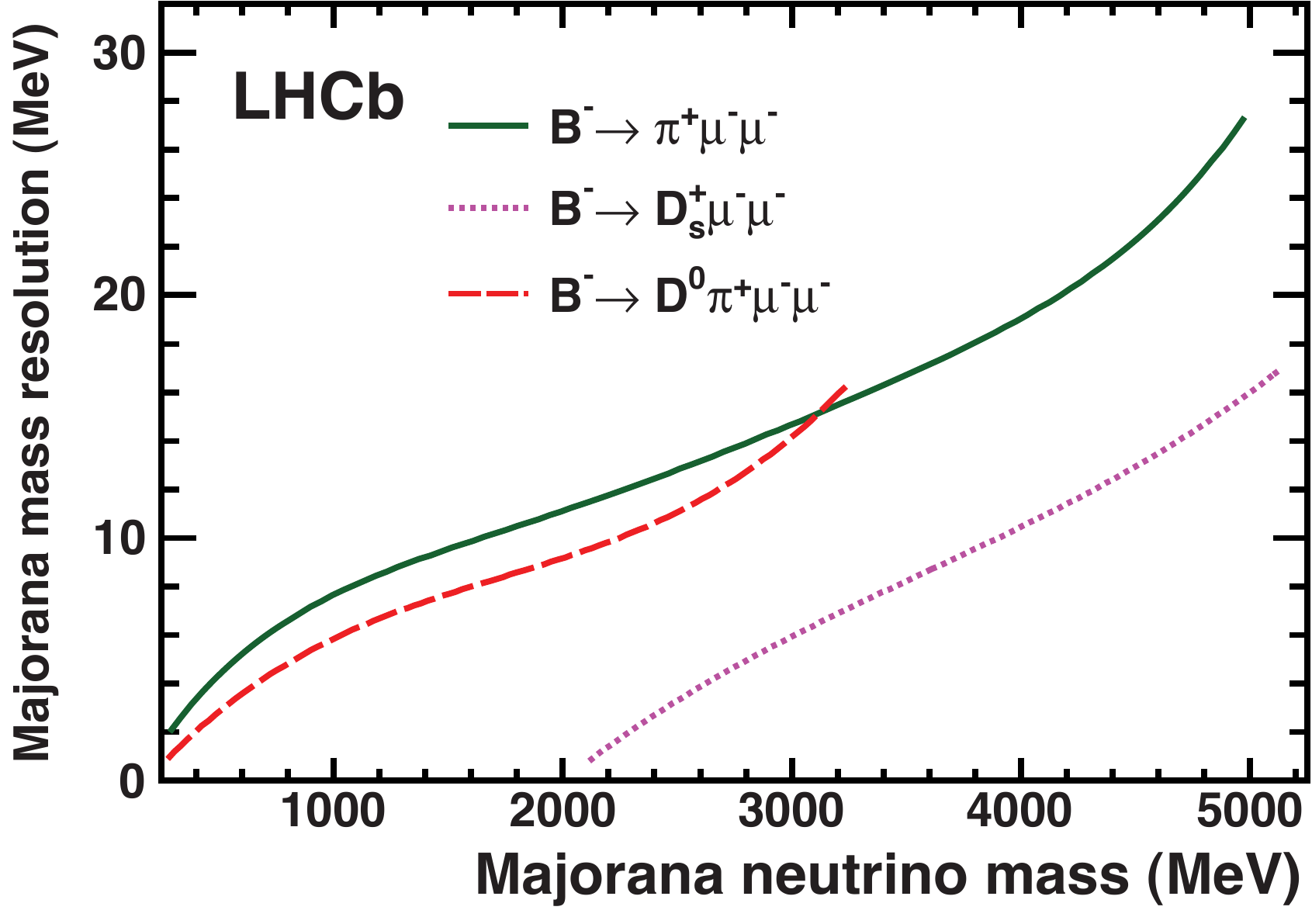}
\caption{Majorana mass resolutions for the three $B^-$ decays as a function of Majorana mass. } \label{Resolutions}
\end{figure}

\begin{figure}[!h]
\centering
\includegraphics[width=4.5in]{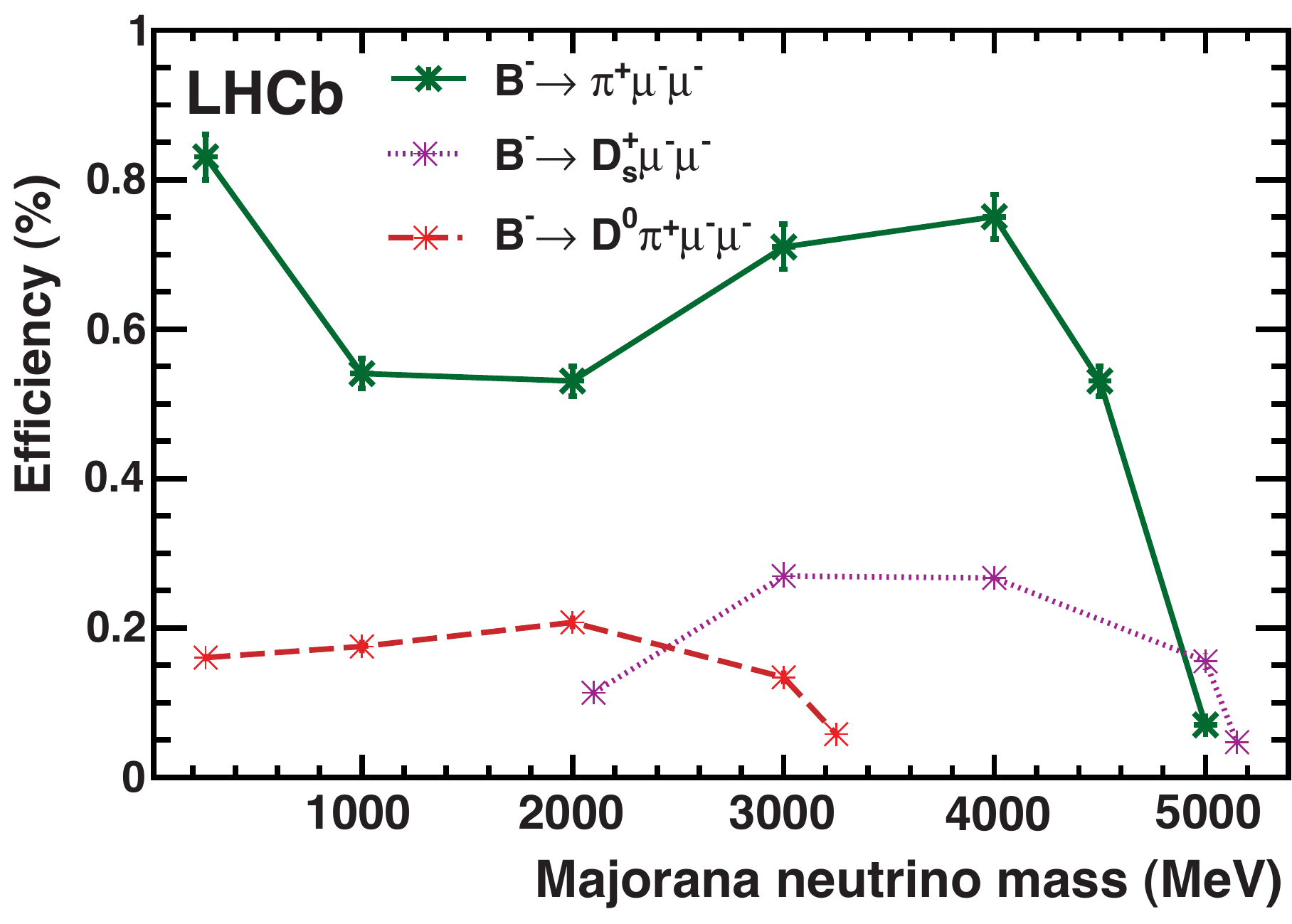}
\caption{Detection efficiencies for the three $B^-$ decays as a function of Majorana mass. Charm meson decay branching fractions are not included.} \label{Overeff}
\end{figure}

Since the putative neutrinos considered here decay into $\pi^+\mu^-$, and are assumed to have very narrow widths, more sensitivity is obtained by examining this mass distribution, shown in  Fig.~\ref{MassSignalN}, for events in the $B^-$ signal region.
 There is no statistically significant signal at any mass. There are three combinations in one mass bin near 2530 MeV; however two of the combinations come from one event, while it is possible to only have one Majorana neutrino per $B^-$ decay.  Upper limits at 95\% confidence level on the existence of a massive Majorana neutrino are set at each $\pi^+\mu^-$ mass by searching a signal region whose width is $\pm3\sigma_N$, where $\sigma_N$ is the mass resolution, at each possible Majorana neutrino mass, $M_N$.  This is done in very small steps in  $\pi^+\mu^-$ mass and so produces a continuous curve.  If a mass combination is found anywhere in the $\pm3\sigma_N$ interval it is considered as part of the observed yield. To set upper limits
the mass resolution and the detection efficiency as a function of  $\pi^+\mu^-$ mass need to be known. Monte Carlo simulation of the mass resolution as a function of  the Majorana neutrino mass is shown in
Fig.~\ref{Resolutions}, along with resolutions of other channels.
The overall efficiencies for different values of  $M_N$ are  shown in Fig.~\ref{Overeff}. A linear interpolation is used to obtain values between the simulated points.

Many systematic errors in the signal yield cancel in the ratio to the normalization channel. The remaining systematic uncertainties are listed in Table~\ref{tab: SysPi}. The largest sources of error are the modeling of the detector efficiency (5.3\%) and the measured branching fractions ${\cal{B}}\left(B^-\to J/\psi K^-\right)$ (3.4\%),  and ${\cal{B}}\left(J/\psi \to\mu^+\mu^-\right)$ (1.0\%).  
\begin{table}[!htb]
\centering
\caption{Systematic uncertainties for $B^- \to \pi^+\mu^- \mu^- $ measurement. }
\label{tab: SysPi}
\begin{tabular}{lcc}
\hline\hline
Selection criteria & Systematic uncertainties (\%)\\\hline 
$K/\pi$ PID & 1.0 \\
$\mu$ PID & 0.6\\
Muon selection & 0.6 \\
Trigger & 1.0 \\
Yields of reference channel & 0.4 \\
Efficiency modeling &  5.3\\
${\cal{B}}(B^- \to J/\psi K^-$) & 3.4\\
${\cal{B}}(J/\psi \to \mu^+ \mu^-$) & 1.0\\\hline
Total  &  6.7\\
\hline
\end{tabular}
\end{table}

\begin{figure}[hbt]
\centering
\includegraphics[width=6.4in]{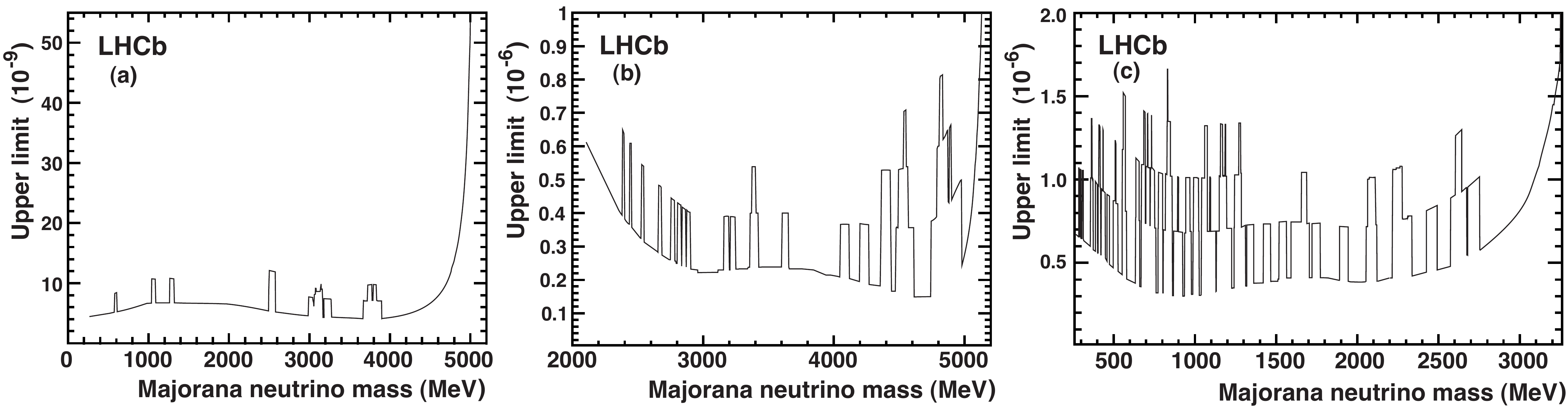}
\caption{Upper limits at 95\%~CL as a function of the putative Majorana neutrino mass,  (a) for ${\cal{B}}(B^- \to \pi^+\mu^- \mu^-)$ as a function of the  $\pi^+\mu^-$ mass, (b)  for ${\cal{B}}(B^-\to D_s^+\mu^-\mu^-$) as a function of the  $D_s^+\mu^-$ mass, and (c) for 
${\cal{B}}(B^-\to D^0\pi^+\mu^-\mu^-)$ as a function of the $\pi^+\mu^-$ mass.} \label{fig: UpperAll}
\end{figure}

To set upper limits on the branching fraction, the number of events $N_{\rm obs}$ at each $M_N$ value (within $\pm3\sigma_N$) are counted, and the procedure described in the last section applied. Estimated background levels are taken from Fig.~\ref{MassSignalN}. Figure~\ref{fig: UpperAll}(a) shows the upper limit on ${\cal{B}}(B^-\to\pi^+\mu^-\mu^-)$ as a function of $M_N$ at 95\%~CL. For most of the neutrino mass region, the limits on the branching ratio are $< 8 \times 10^{-9}$. Assuming a phase space decay of the $B^-$ we also determine
\begin{equation}
{\cal{B}}(B^-\to\pi^+\mu^-\mu^-)<1.3 \times 10^{-8}~{\rm at~95\%~CL.}\nonumber
\end{equation}
These limits improve on previous results from CLEO $({<1.4\times 10^{-6}}~{\rm at~90\%~CL})$ \cite{Edwards:2002kq}, and LHCb  $({<5.8\times 10^{-8}}~{\rm at~95\%~CL})$ \cite{Aaij:2011ex}.

\section{\boldmath Analysis of $B^-\to D_s^+\mu^-\mu^-$}
\label{sec:Dsmumu}
\begin{figure}[hbt]
\centering
\includegraphics[width=4.5in]{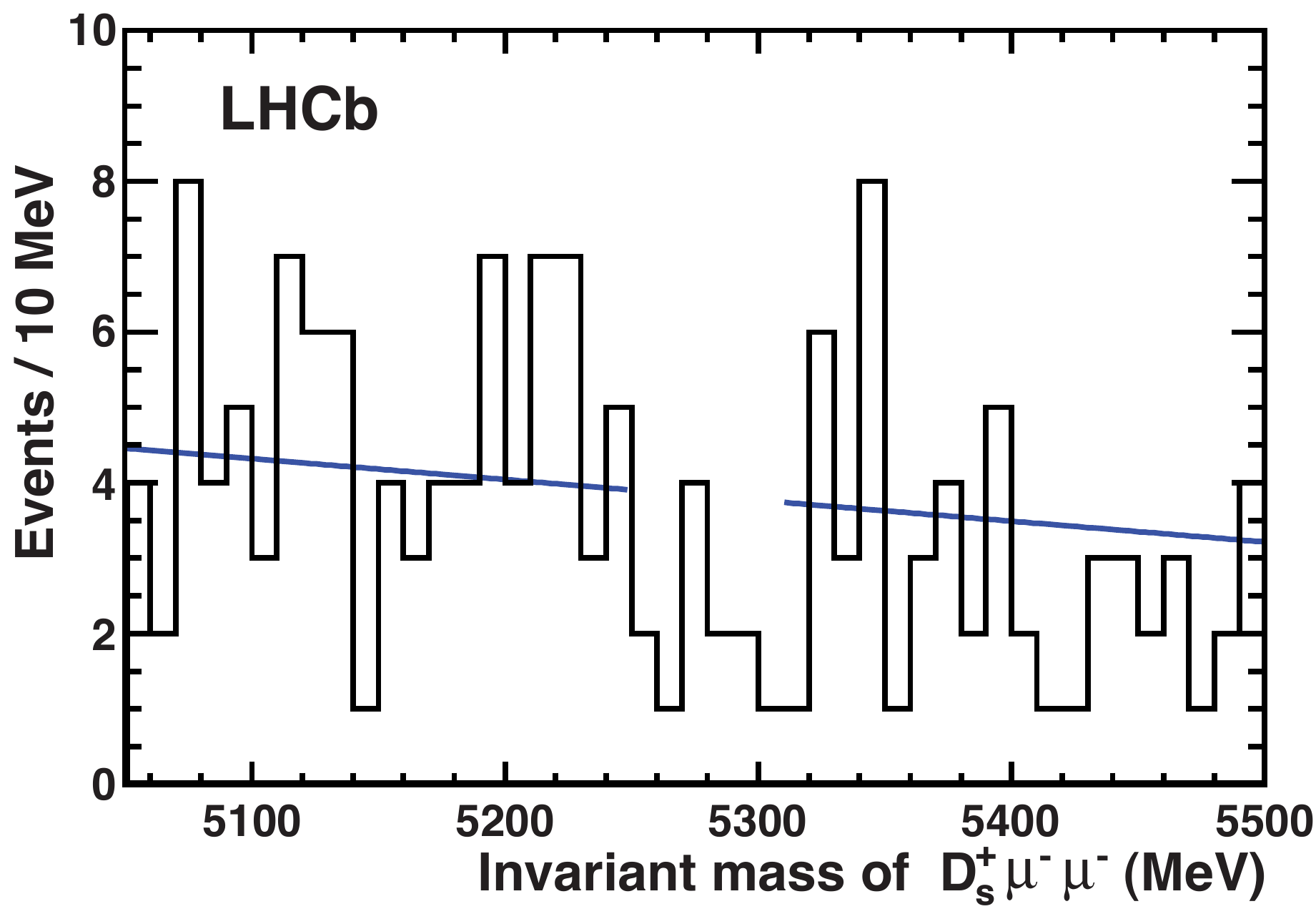}
\caption{Invariant mass spectrum for $B^-\to D_s^+\mu^-\mu^-$ candidates. The line shows the fit to the data excluding
the $B^-$ mass signal region.  } \label{MassFitDs}
\end{figure}

The process $B^-\to D_s^+\mu^-\mu^-$ is similar to $B^-\to \pi^+\mu^-\mu^-$, with the difference being that the heavy neutrino can decay into $D_s^+\mu^-$. Here we consider only $D_s^+\to K^+K^-\pi^+$ decays. 
Our analysis follows a similar procedure used for the $\pi^+\mu^-\mu^-$ channel. Candidate $D_s^+\to  K^+K^-\pi^+$ decays are selected by having an invariant mass within $\pm$25 MeV of the $D_s^+$ mass. A Majorana neutrino candidate decay is then looked for by by having the $D_s^+$ candidate decay tracks form a vertex with an opposite-sign muon candidate. Then this neutrino candidate must form a vertex with another muon of like-sign to the first one consistent with a $B^-$ decay detached from the primary vertex. The invariant mass spectrum of $D_s^+\mu^-\mu^-$ candidates is shown in Fig.~\ref{MassFitDs}. The mass resolution is 15.5$\pm$0.3 MeV.

There are 12 events within the $B^-$ candidate mass region; it appears that there is a dip in the number of events here. An unbinned fit to the data in the sidebands gives an estimate of 22 events. The fluctuation at the $B^-$ mass, therefore, is about two standard deviations. Peaking background contributions at the level of current sensitivity are negligible ($\sim$3$\times 10^{-4}$); thus only combinatorial background is considered.

\begin{figure}[hbt]
\centering
\includegraphics[width=4.5in]{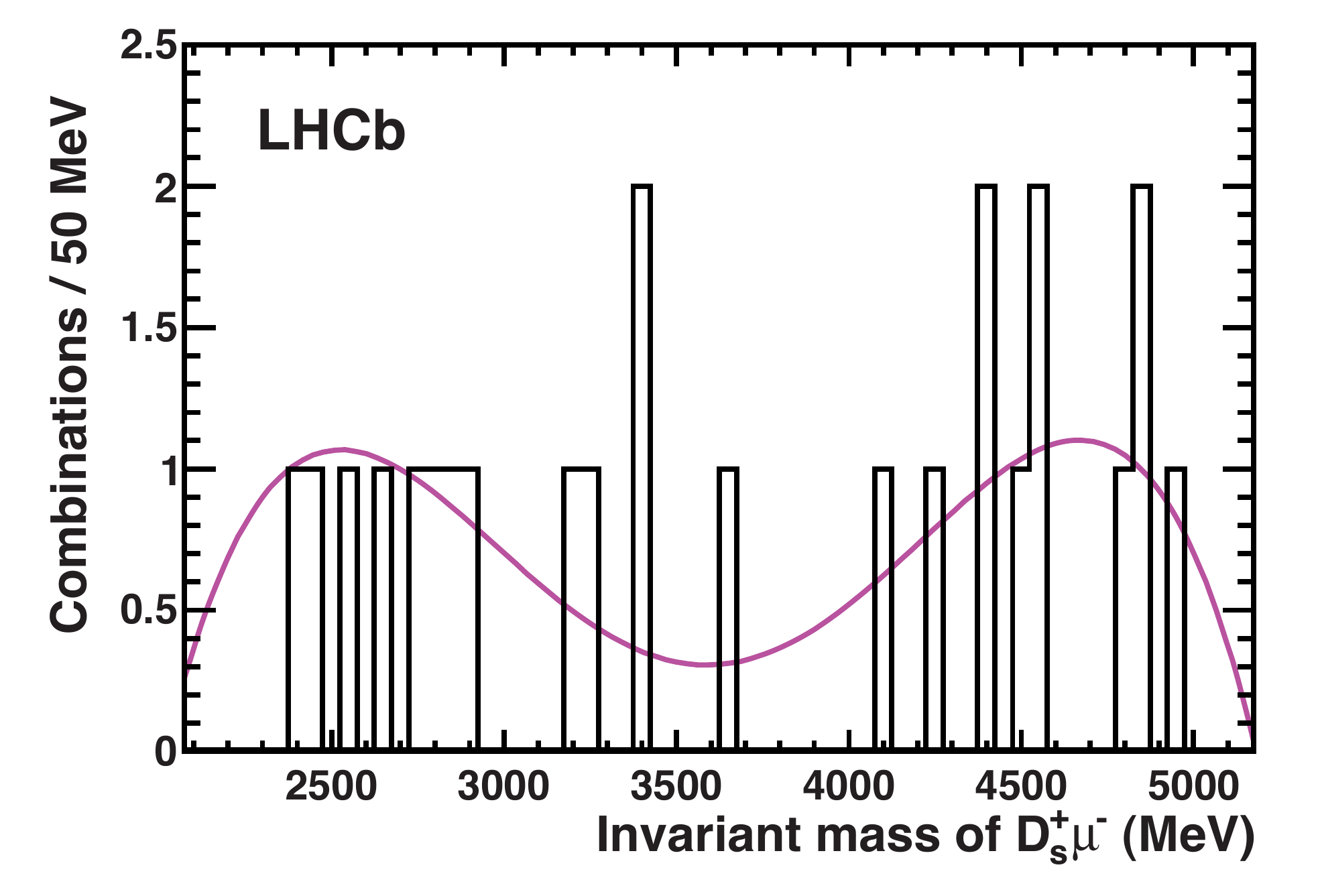}
\caption{Invariant mass spectrum of $D_s^+\mu^-$ from $B^-\to D_s^+\mu^-\mu^-$ events in the signal region with the background estimate superimposed (solid curve). There are two combinations per event. } \label{DsEstimated}
\end{figure}

After selecting the events in the $B^-$ signal region, we plot the $D_s^+\mu^-$ invariant mass distribution, which is shown in Fig.~\ref{DsEstimated}.
A background estimate is made using the sideband data in $B^-$ candidate mass (see Fig.~\ref{MassFitDs}), by fitting to a 4th order polynomial. The background estimated from the sidebands is also shown in the figure. The normalization is absolute and in agreement with the data.  The data in the signal region is consistent with the background estimate. The systematic error due to the fitting procedure is estimated using the difference between this fit and the one obtained using a 6th order polynomial. 

The overall efficiencies for different values of  $M_N$ are shown in Fig.~\ref{Overeff}. As done previously,
during the scan over the accessible Majorana neutrino mass region we use a $\pm$3$\sigma_N$ mass window around a given Majorana mass. The resolution is plotted in
Fig.~\ref{Resolutions} as a function of $M_N$.
Systematic uncertainties are listed in
Table~\ref{tab: SysD}.

Again we provide upper limits as a function of the Majorana neutrino mass, shown in Fig.~\ref{fig: UpperAll}(b), only taking into account combinatorial background in this case as the peaking background is absent. 
For neutrino masses below 5 GeV, the limits on the mass dependent branching fractions are mostly $<6 \times 10^{-7}$. 
We also determine an upper limit on the total branching fraction. Since the background estimate of 22 events exceeds the observed level of 12 events we use the $CL_s$ method for calculating the upper limit \cite{Read:2002hq}. Assuming a phase space decay of the $B^-$ we find
\begin{equation}
{\cal{B}}(B^-\to D_s^+\mu^-\mu^-)< 5.8\times 10^{-7}~{\rm at~95\%~CL.}\nonumber
\end{equation}

\section{\boldmath Analysis of $B^-\to D^0\pi^+\mu^-\mu^-$}
\begin{figure}[hbt]
\centering
\includegraphics[width=4.5in]{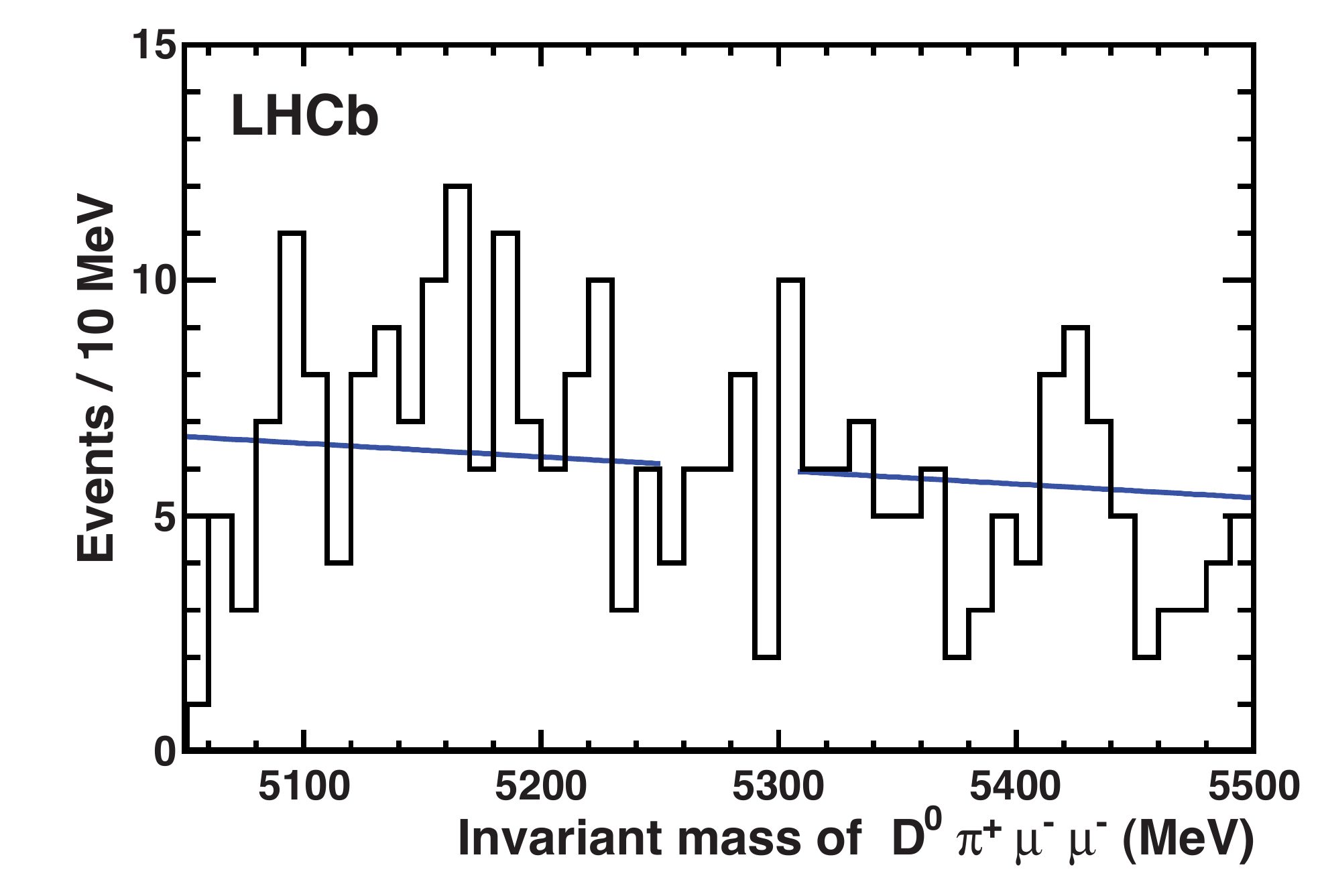}
\caption{Invariant mass distribution of $D^0\pi^+ \mu^- \mu^- $. The solid line shows a linear fit to the data in the sidebands of the $B^-$ signal region.}
\label{fig: MassFitD0}
\end{figure}

A prolific source of neutrinos is semileptonic $B^-$ decay. Majorana neutrinos could be produced via semileptonic decays as shown in Fig.~\ref{B-Majorana}(c). Here the mass range probed is smaller than in the case of $\pi^+\mu^-\mu^-$ due to the presence of the $D^0$ meson in the final state. The sensitivity of the search in this channel is also limited by the need to reconstruct the $D^0\to K^-\pi^+$ decay. 
We do not explicitly veto $D^{*+}\to \pi^+D^0$ decays as this would introduce an additional systematic uncertainty.
The invariant mass distribution of $D^0 \pi^+ \mu^- \mu^-$ is shown in Fig.~\ref{fig: MassFitD0}. The mass resolution is 14.4$\pm$0.2 MeV.

Peaking backgrounds are essentially absent; the largest source is $B^- \to D^0 \pi^- \pi^- \pi^+$ which contributes only 0.13 events in the signal region.  The combinatorial background, determined by a linear fit to the sidebands of the $B^-$ signal region, predicts 35.9 events, while the number observed is 33. 
 
\begin{figure}[hbt]
\centering
\includegraphics[width=4.5in]{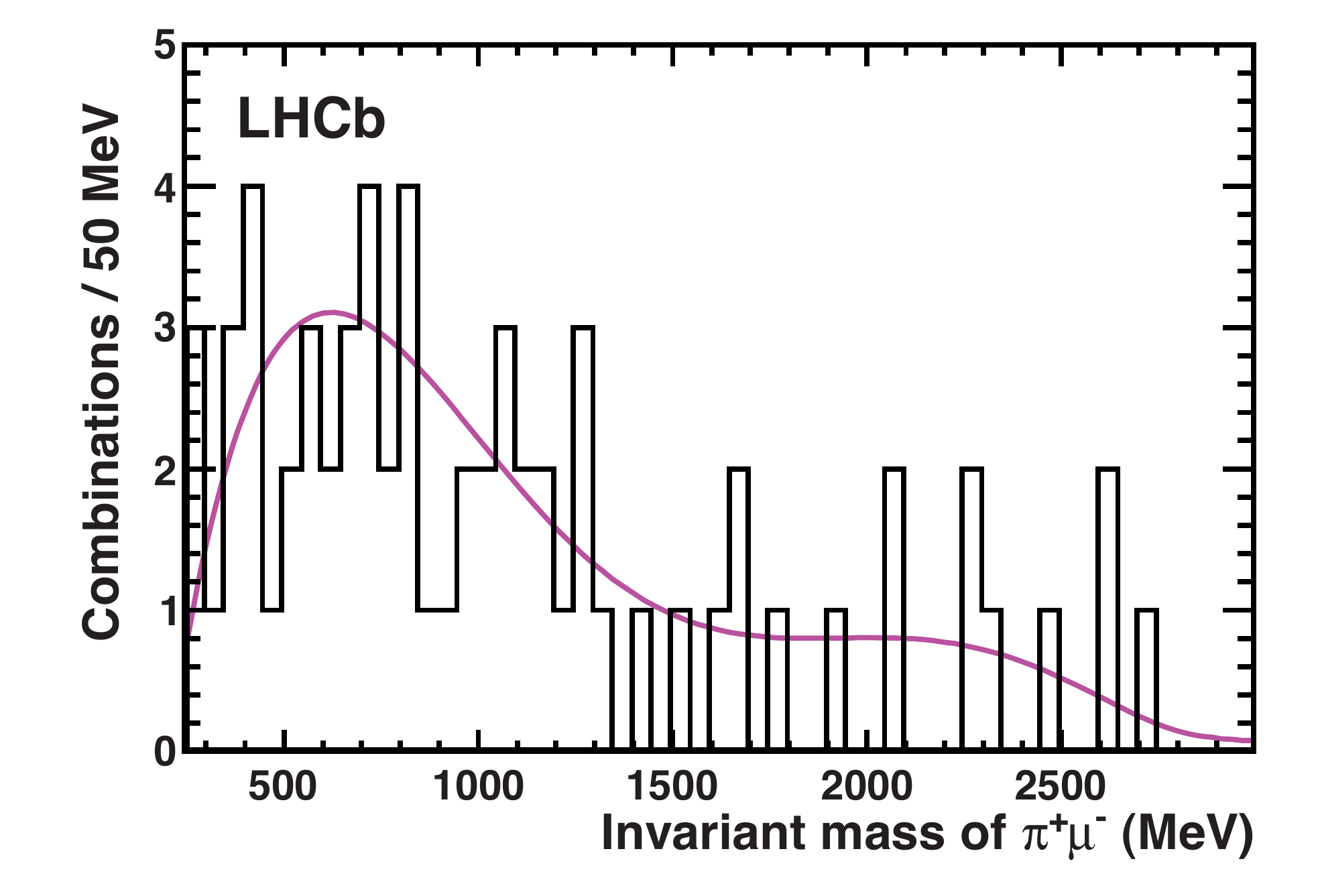}
\caption{Invariant mass distribution of $\pi^+\mu^-$ for $B^-\to D^0\mu^-\mu^- \pi^+$ in the signal region and with estimated background distribution superimposed.  There are two combinations per event.}
\label{D0Estimated}
\end{figure}

The $\pi^+\mu^-$ invariant mass for events within two standard deviations of the $B^-$ mass is shown in Fig.~\ref{D0Estimated}. The background shape is estimated by a 5th order polynomial fit to the sideband data (see Fig.~\ref{fig: MassFitD0}) and also shown on the figure. The systematic error on this background is estimated using a 7th order polynomial fit.

The $\pi^+\mu^-$ mass resolution is 
shown in Fig.~\ref{Resolutions}. The $M_N$ dependent efficiencies are shown in Fig.~\ref{Overeff}. They vary from 0.2\% to 0.1\% over most of the mass range.
Systematic errors are listed in Table~\ref{tab: SysD}. The largest sources of error are the trigger, and the $M_N$ dependent efficiencies.

The upper limits for ${\cal{B}}(B^-\to D^0\pi^+\mu^-\mu^-)$ as a function of the $\pi^+\mu^-$ mass are shown in Fig.~\ref{fig: UpperAll}(c). 
For Majorana neutrino masses $<$ 3.0 GeV, the upper limits are less than $1.6 \times 10^{-6}$ at 95\% CL.
The limit on the branching fraction assuming a phase space decay is
\begin{equation}
{\cal{B}}(B^-\to D^0 \pi^+\mu^-\mu^-)<1.5 \times 10^{-6}~{\rm at~95\%~CL.}\nonumber
\end{equation}


\section{Conclusions}

A search has been performed for Majorana neutrinos in the $B^-$ decay channels, $D^{(*)+} \mu^- \mu^-$, $\pi^+ \mu^-\mu^-$, 
$D_s^+ \mu^- \mu^-$, and  $D^0\pi^+ \mu^- \mu^- $ that has only yielded upper limits.
The $D^{(*)+}\mu^-\mu^-$ channels may proceed via virtual Majorana neutrino exchange and thus are sensitive to all Majorana neutrino masses.  They also could occur via the same annihilation process as the other modes, though this would be Cabibbo suppressed. The other channels provide limits for neutrino masses between 260 and 5000 MeV.
The bounds are summarized in Table~\ref{tab:Sum}.
These limits are the most restrictive to date.

\begin{table}[!htb]
\centering
\caption{Summary of upper limits on branching fractions. Both the limits on the overall branching fraction assuming a phase space decay, and the range of limits on the branching fraction as a function of Majorana neutrino mass ($M_N$) are given. All limits are at 95\%~CL.}
\label{tab:Sum}
\begin{tabular}{lccccccc}
\hline\hline
Mode    &                    ${\cal{B}}$ upper limit  &  Approx. limits as function of $M_N$ \\\hline
$D^+\mu^-\mu^-$ & $6.9\times 10^{-7}$ &\\
$D^{*+}\mu^-\mu^-$ & $ 2.4\times 10^{-6}$ &\\
$\pi^+\mu^-\mu^-$ &  $1.3\times 10^{-8}$ & $(0.4-1.0)\times 10^{-8}$ \\
$D_s^+\mu^-\mu^-$ & $5.8\times 10^{-7}$ & $(1.5 - 8.0)\times 10^{-7}$\\
$D^0\pi^+\mu^-\mu^-$ &  $1.5\times 10^{-6}$ &$(0.3-1.5)\times 10^{-6}$\\
\hline\hline
\end{tabular}
\end{table}

\begin{figure}[hbt]
\centering
\hspace*{-4.0mm}\includegraphics[width=6.4in]{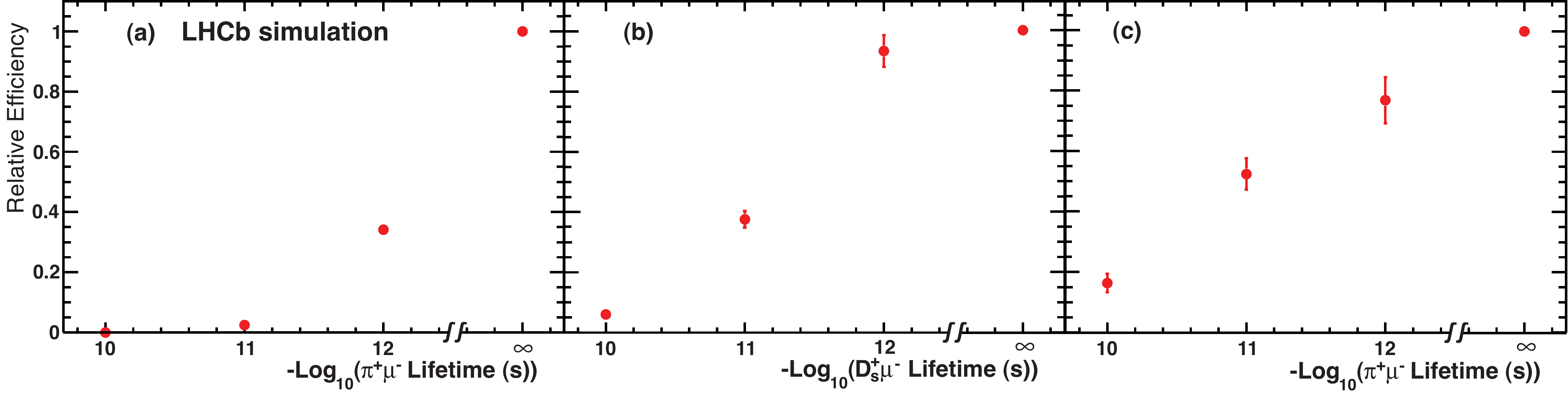}
\caption{Relative efficiencies as a function of Majorana neutrino lifetime for (a) $B^-\to \pi^+\mu^-\mu^-$ for a mass of 3 GeV, (b)  $B^-\to D_s^+\mu^-\mu^-$ for a mass of 3 GeV, and (c)  $B^-\to D^0\pi^+\mu^-\mu^-$ for a Majorana neutrino mass of 2 GeV. Where the error bars are not visible, they are smaller than radii of the points.
} \label{fig: LifetimePi}
\end{figure}

Our search has thus far ignored the possibility of a finite neutrino lifetime. Figure~\ref{fig: LifetimePi} shows the relative detection efficiency as a function of Majorana neutrino lifetime, for (a) $B^-\to \pi^+\mu^-\mu^-$ for a mass of 3 GeV, (b)  $B^-\to D_s^+\mu^-\mu^-$ for a mass of 3 GeV, and (c)  $B^-\to D^0\pi^+\mu^-\mu^-$ for a mass of 2 GeV. All sensitivity is lost for lifetimes longer than $10^{-10}$\,s to $10^{-11}$\,s, depending on the decay mode. Note that for the $D^{(*)+}\mu^-\mu^-$ final states the detection efficiency is independent of the neutrino lifetime, since the neutrino acts a virtual particle.

\begin{figure}[hbt]
\centering
\includegraphics[width=4.5in]{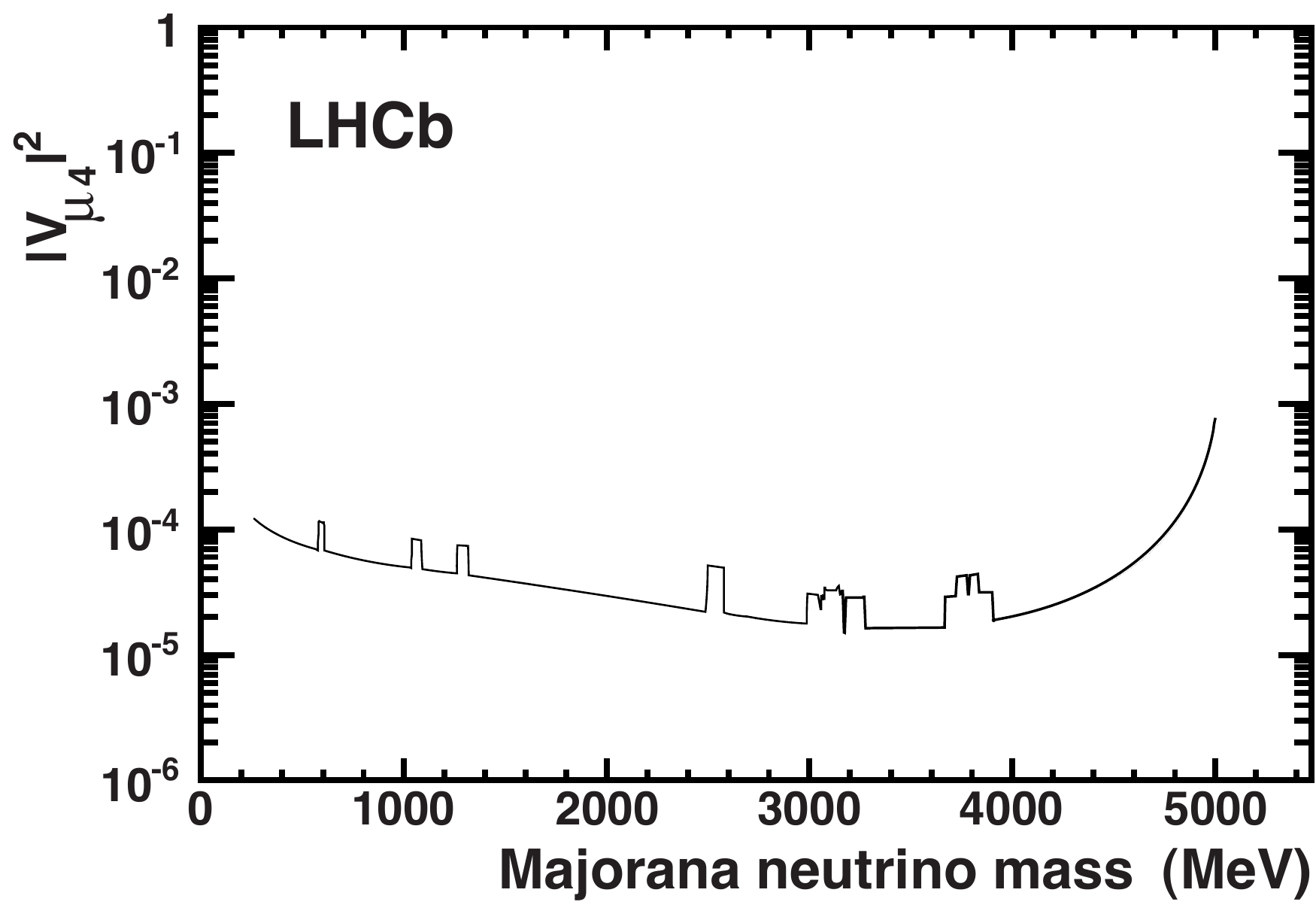}
\caption{Upper limits on $|V_{\mu4}|^2$ at 95\% CL as a function of the Majorana neutrino mass from the $B^- \to \pi^+ \mu^- \mu^-$ channel.} \label{fig: ConstrainPi}
\end{figure}

Our upper limits in the $\pi^+ \mu^- \mu^-$ final state can be used to establish neutrino mass dependent upper limits on the coupling  $|V_{\mu4}|$ of a heavy Majorana neutrino to a muon and a virtual $W$.
The matrix element has been calculated in Ref.~\cite{Atre:2009rg}.
The results are shown in Fig.~\ref{fig: ConstrainPi} as a function of $M_{N}$. 
A model dependent calculation of ${\cal{B}}(B^- \to D^0 \pi^+ \mu^- \mu^-)$ can also be used to extract $|V_{\mu4}|$ \cite{Quintero:2011yh}, but the $\pi^+\mu^-\mu^-$ mode is more sensitive.
For the $D^{(*)+}\mu^-\mu^-$ channels upper limits cannot be extracted until there is a theoretical calculation of the hadronic form-factor similar to those available for neutrinoless double $\beta$ decay.

\section*{Acknowledgements}

\noindent We express our gratitude to our colleagues in the CERN accelerator
departments for the excellent performance of the LHC. We thank the
technical and administrative staff at CERN and at the LHCb institutes,
and acknowledge support from the National Agencies: CAPES, CNPq,
FAPERJ and FINEP (Brazil); CERN; NSFC (China); CNRS/IN2P3 (France);
BMBF, DFG, HGF and MPG (Germany); SFI (Ireland); INFN (Italy); FOM and
NWO (The Netherlands); SCSR (Poland); ANCS (Romania); MinES of Russia and
Rosatom (Russia); MICINN, XuntaGal and GENCAT (Spain); SNSF and SER
(Switzerland); NAS Ukraine (Ukraine); STFC (United Kingdom); NSF
(USA). We also acknowledge the support received from the ERC under FP7
and the Region Auvergne.

\clearpage
\newpage
\ifx\mcitethebibliography\mciteundefinedmacro
\PackageError{LHCb.bst}{mciteplus.sty has not been loaded}
{This bibstyle requires the use of the mciteplus package.}\fi
\providecommand{\href}[2]{#2}

\end{document}